\documentstyle[11pt,aaspp4,colordvi,epsf]{article}

\newcommand{\kms}{km\, s$^{-1}$}
\newcommand{\Msun}{$M_{\odot}$}

\newcommand{\fesc}{$f_{esc}$}

\input epsf

\begin{document}

\title{ZERO-METALLICITY STARS AND THE EFFECTS OF THE \\ FIRST 
STARS ON REIONIZATION} 
\author{JASON TUMLINSON and J. MICHAEL SHULL\altaffilmark{1}}
\affil{Center for Astrophysics and Space Astronomy, \\ 
Department of Astrophysical and Planetary Sciences, \\ 
University of Colorado, CB 389, Boulder, CO, 80309 \\ Electronic Mail:
(tumlinso, mshull)@casa.colorado.edu} \altaffiltext{1}{Also at JILA,
University of Colorado and National Institute of Standards
      and Technology.} 

\begin{abstract}
We present stellar structure and atmosphere models of metal-free stars
and examine them from a cosmological point of view.  Metal-free stars
exhibit high effective temperatures and small sizes relative to
metal-enriched stars of equal mass.  These unique physical
characteristics enhance the ionizing photon production by metal-free
stars, particularly in the \ion{He}{2} ($h\nu \geq 4$ Ryd) continuum.
The star formation rate of metal-free stars necessary to reionize the
hydrogen in the universe by $z = 5$ is consistent with the inferred
star formation rate at that epoch.  However, the hard stellar spectra
are inconsistent with the observations of \ion{He}{2} opacity in the
IGM at $z \sim 3$, indicating that the period of metal-free star
formation ended before that epoch.  We examine the effects of these
stars on the ionization balance of the IGM, the radiative feedback of
the first luminous objects, and the extragalactic radiation field. We
comment on the prospects for detecting metal-free stellar populations
with the $\lambda1640$ and $\lambda4686$ recombination lines of
\ion{He}{2}.
\end{abstract}


\section{INTRODUCTION}

We know from observational constraints that the hydrogen in the
intergalactic medium (IGM) was reionized by redshift $z \sim 5$
(Schneider, Schmidt, \& Gunn~1991) and that the \ion{He}{2} was
reionized at $z \sim 3$ (Reimers et al.~1997).  However, the exact
nature of the ionizing sources is still uncertain. The observed drop in
the space density of bright quasars at $z \geq 3$ (Pei~1995) suggests
that early stellar populations played a role in the reionization of
hydrogen, but it is not known whether hot stars produced photons at
rates sufficient to ionize the universe before this epoch.

Our understanding of reionization is closely connected to our knowledge
of the extragalactic radiation field.  Models of reionization assume an
extragalactic spectrum composed of the distinct intrinsic spectra of
active galactic nuclei (AGN) and star-forming galaxies. This composite
spectrum determines the reionization epoch, the \ion{He}{2}/\ion{H}{1}
ionization ratio, and metal-line absorption ratios at $z < 5$.  Models
by Giroux \& Shull (1997) of the observed \ion{Si}{4}/\ion{C}{4} ratio
at $z \sim 3$ (Songaila \& Cowie~1996) and subsequent work by Haardt \&
Madau (1996) and Fardal et al.~(1998) on the \ion{He}{2} Gunn-Peterson
effect demonstrated that the observations are consistent with an
extragalactic spectrum produced by a mixture of QSOs and hot stars.
These conclusions depend, however, on the assumed shape of the ionizing
spectrum of stars. While the details of these spectra vary (Sutherland
\& Shull~1999; Leitherer et al.~1999), one common element is the
assumption that hot stars contribute few \ion{He}{2} ionizing photons.

Some models of reionization assume a phenomenological prescription for
star formation, in which a single parameter describes the conversion
efficiency of mass to stars and then to ionizing photons (Gnedin \&
Ostriker~1997). Others use existing model grids of stellar structure
and atmospheres designed for application to metal-poor environments
(Haiman \& Loeb~1997).  The first method ignores the details of star
formation, ionizing photon production, and radiation escape from the
immediate regions of star formation. The second method has the drawback
of applying theoretical calculations of metal-poor stars to the very
different regime of zero metallicity.  The existing grids of stellar
evolution tracks extend to $Z = 0.001$ (Schaerer et al.~1996 and
references therein).  Existing model atmospheres extend to $Z = 2
\times 10^{-7}$ but are limited in the range of stellar parameters
(Kurucz~1992).  These models are meant for application to
low-metallicity starbursts (Leitherer et al.~1999) and metal-poor
galactic halo populations.  As shown by existing models of metal-free
stars (Ezer 1972; Ezer \& Cameron 1971; El Eid et al. 1983), however,
stars with $Z \sim 0.001$ are quite different from their $Z = 0$
counterparts.  Thus, when we consider the effects of stellar
populations on reionization, we must use true metal-free models to
predict the ionizing photon production of the first generation of
stars.

In this {\em Letter} we adopt the common term ``Population III'' or
``Pop~III'' for metal-free stars, which are understood to have formed
from primeval gas.  Although extremely metal-poor populations ($Z
\lesssim 0.001$) may fit an observer's definition of Pop~III, we apply
that label to metal-free stars only.  In \S~2 we present structure and
atmosphere models of metal-free stars, and in \S~3 we predict ionizing
photon yields from these model stars.  In \S~4 we evaluate the effects
of these models on the epoch of reionization, and in \S~5 we comment on
further cosmological implications of metal-free stellar populations.

\section{STRUCTURE AND ATMOSPHERE MODELS} 

The models presented here are static stellar structure models
calculated using a fitting-method technique that incorporates OPAL
radiative opacities (Rogers \& Iglesias 1992) and analytic expressions
for energy generation.  These models were used to predict the effective
temperature ($T_{\rm eff}$), luminosity ($L$), and surface gravity
($g$) of stars with mass 2 -- 90 \Msun\ (at 5 -- 10 \Msun\ intervals).
There is currently no full set of evolutionary tracks for
metal-free stars, and our models do not incorporate evolution.
However, the existing evolutionary tracks for metal-free
stars (Castellani, Chieffi, \& Tornambe~1983; Chieffi \& Tornambe~1984)
show that, like their metal-enriched counterparts, these stars become
systematically cooler, larger, and more luminous over their H-burning
lifetimes, which are similar in duration.  Therefore, we assume that
metal-free tracks differ in their first-order characteristics only in
their starting point on the Hertzsprung-Russell (HR) diagram.  If so,
the ``gain'' in ionizing photons at $Z = 0$ is maintained throughout
the main sequence (MS) lifetime of the star, when most of its
ionizing radiation is released.

Existing evolutionary tracks of metal-free stars show that these stars
may build up a small fraction of C nuclei ($Z_{C} \sim 10^{-10}$) via
the triple-$\alpha$ process before they join the H-burning main
sequence (El Eid et al. 1983; Castellani et al.~1983).  Following this
result, we assume that stars with $M \geq 15$ \Msun\ are enriched 
to $Z_{\rm C} = 10^{-10}$ via triple-$\alpha$ burning prior to the
onset of MS H-burning. We use these pre-enriched models in all
our analysis.  We will explore the detailed effects of pre-MS
self-enrichment in a later paper, once a comprehensive grid of tracks
is available.

Figure~\ref{fig1} shows an HR diagram for zero-age main sequence models
with Pop I and Pop~III metallicities.  The most striking feature of the
metal-free models is the high temperature they maintain at their
photospheres.  These stars derive their nuclear energy from a
combination of inefficient proton-proton burning and CNO burning with
the small fraction of C built up in the pre-MS phase (Castellani et
al.~1983; El Eid et al.~1983).  As a result of lower energy generation
rates in the convective core, they maintain core temperatures in excess
of 10$^{8}$~K to support the mass against gravitational collapse.
Together with reduced radiative opacity in their envelopes, the high
core temperatures of Pop~III stars make these stars hotter and smaller
than their metal-enriched counterparts.  Using homology relations, and
assuming constant (electron scattering) opacity and a CNO burning rate
$\epsilon \propto \rho T_c^{12} (Z_{\rm CN}/Z_{\odot})$, we find that
massive stars have $R \propto (Z_{\rm CN}/Z_{\odot})^{1/12}$ and
$T_{\rm eff} \propto (Z_{\rm CN}/Z_{\odot})^{-1/24}$, in good agreement
with our numerical models.

With $g$ and $T_{\rm eff}$ in hand, we need a model atmosphere to
derive the spectral luminosity distribution $L_{\nu}$.  A model
atmosphere is necessary because a simple blackbody curve for each
$T_{\rm eff}$ will not accurately reproduce the spectrum near the
ionization edges of \ion{H}{1}, \ion{He}{1}, and \ion{He}{2}.  For each
stellar model and $\log g$ -- $T_{\rm eff}$ pair, we calculated a
static, non-LTE model atmosphere that was used to predict the spectral
luminosity distribution for the star. We used the atmosphere code
TLUSTY (Hubeny \& Lanz~1995) to produce the model atmospheres and its
included package SYNSPEC to produce the continuum spectra.
Figure~\ref{fig2} illustrates the change in the spectral distribution
of a 15 \Msun\ star at $Z = 0.001$ ($T_{\rm eff} = $ 36,000~K) and $Z =
0$ ($T_{\rm eff} = $ 63,000~K).  The key between the two spectra 
if the high $T_{\rm eff}$ of the $Z = 0$ model. A Pop II star of the same 
$T_{\rm eff}$ would exhibit a continuum shape similar to the Pop~III model  
but attenuated by metal-line absorption. However, a Pop II star at 15 \Msun\ 
is unlikely to reach $T_{\rm eff} = $ 63,000 K during its 
MS lifetime. 

\section{IONIZING PHOTON YIELDS}

The model atmospheres were used to predict the rates of ionizing 
photon production, $Q_{i}$, in photons s$^{-1}$, for all the modeled stars: 
\begin{eqnarray}
Q _{i} = 4 \pi R^{2}_{*} \int _{\nu _{i}} ^{\infty} {\frac{F
_{\nu}}{h\nu} \; d\nu }, 
\end{eqnarray}
where $R_{*}$ is the radius of the star, $F_{\nu}$ is the spectral flux
distribution in erg cm$^{-2}$ s$^{-1}$ Hz$^{-1}$, and the indices $i$ =
0, 1 and 2 correspond to integration over the \ion{H}{1}, \ion{He}{1}, 
and \ion{He}{2} ionizing continua, $h\nu \geq h\nu_{i}$ (13.60, 24.58,
and 54.40 eV, respectively).  The $Q_{i}$ for Pop~III stars appear in
Figure~\ref{fig3}.  Individual metal-free stars emit far more of their
energy in photons with $h\nu > 13.6$ eV than do their Pop~I and
Pop~II counterparts.  This increase produces a 50\% gain in the total
ionizing photon production at high mass (40 -- 70 \Msun) and gains by
factors of $2 - 40$ at moderate mass (10 -- 30 \Msun). Above 70 \Msun,
a larger fraction of the star's energy is released above 1 Ryd, but the
average photon energy increases such that the overall gain in $Q_{0}$
is modest.

A striking feature of the ionizing photon production is the high
fraction of the photons emitted with energy sufficient to ionize
\ion{He}{1} and \ion{He}{2}.  These fractions are expressed by the
ratios $Q_{1}/Q_{0}$ and $Q_{2}/Q_{0}$, displayed in
Figure~\ref{fig3}.  High-mass Pop~III stars emit 60 -- 70\% of their
ionizing photons in the \ion{He}{1} continuum and up to 12\% of these
photons in the \ion{He}{2} continuum.  For comparison, $Q_{1}/Q_{0}
\simeq$ 0.2 -- 0.4 for Pop~I O3 -- O5 stars with $T_{\rm eff}$ in the range
45,000 -- 51,000 K (Vacca, Garmany, \& Shull~1996; Schaerer \& de
Koter~1997).

The overall gain in integrated ionizing photon production is best
evaluated with synthetic spectra of stellar clusters (Sutherland \&
Shull~1999; Leitherer et al.~1999), which compare the rates $Q_{i}$ per
unit mass of stellar material.  Synthetic spectra of Pop~III and Pop~II
zero-age clusters with a standard initial mass function (Salpeter IMF
with $0.1 \leq M / M_{\odot} \leq 100$) appear in Figure~\ref{fig4}.
The gain in $Q_{0}$ (s$^{-1}$) is near 50\% for this IMF. However,
$\log Q_{1}$ increases from 52.4 to 52.9, and $\log Q_{2}$ increases
from 46.1 to 51.7.  This dramatic increase in the capability of stars
to ionize \ion{He}{1} and \ion{He}{2} are not predicted by populations
that approximate Pop~III with existing metal-poor models.

\section{IMPLICATIONS FOR REIONIZATION}

Metal-free stars emit 50\% more ionizing radiation per unit mass of
stellar material than metal-enriched stars.  Given the uncertain
efficiency of star formation out of primeval gas (Abel et al.~1998),
our conclusion that Pop~III stars are more efficient ionizing sources
(per unit mass) does not necessarily mean that they are more capable
than other stellar populations of reionizing the universe.  For this
reason, we calculate a star formation rate (SFR) for metal-free stars
necessary to reionize the universe and test this quantity against
observations.

Madau, Haardt, \& Rees~(1999) imposed the condition that the universe
is reionized when the number of ionizing photons emitted in one
recombination time equals the mean number of hydrogen atoms.
Accounting for clumping of the IGM and assuming a typical rate for
ionizing photon production per unit mass, they found that the critical
SFR required to reionize by $z = 5$ is 0.013 $f_{esc}^{-1}$
\Msun\ yr$^{-1}$ Mpc$^{-3}$, where \fesc\ is the fraction of ionizing
photons that escape into the IGM (see Tumlinson et al.~1999 for a
review).  Based on the per-mass photon emission rates of the Pop~III
cluster in Figure~\ref{fig4}, we estimate the critical rate of
metal-free star formation to be $0.008 f_{esc}^{-1}$ \Msun\ yr$^{-1}$
Mpc$^{-3}$. This requirement would be comparable to the inferred
(highly uncertain) SFR at $z = 5$ if \fesc\ = 0.40
(Madau, Mozetti, \& Dickenson~1998).

In addition to enhanced ionizing capability, Pop~III stars
convert less of their initial mass into metals. Preliminary evaluation
of the metal yields of Pop~III stars (Woosley \& Weaver~1995) indicates
that for every solar mass of stars formed, 0.007 \Msun\ in metals are
released. Assuming that this value holds throughout the Pop~III epoch,
then the critical SFR for H reionization by Pop~III
stars implies a metal enrichment rate of $10^{-4}$ \Msun\ yr$^{-1}$
Mpc$^{-3}$.  Over the 500 Myr between $z = 10$ and $z = 5$, Pop~III
stars would enrich the universe to a mean metallicity $\langle Z
\rangle \sim 6 \times 10^{-4}\, Z_{\odot}$ ($H_{0} = 50$
\kms\ Mpc$^{-1}$, $\Omega_{b} = 0.08$), 20\% of the minimum value $Z =
10^{-2.5}\, Z_{\odot}$ observed at $z \sim 3$ and $\sim$2\% of the
average metallicity in damped Ly$\alpha$ systems at that redshift
(Pettini~1999).

Our new models of metal-free stars raise the possibility that stars are
also responsible for \ion{He}{2} reionization.  Various studies
(Reimers et al.~1997; Heap et al.~1999; Hogan, Anderson, \&
Rugers~1997) constrain the epoch of \ion{He}{2} reionization to $z \sim
3$ with observations of patchy \ion{He}{2} absorption.  Conventional
wisdom (Fardal et al.~1998; Madau et al.~1999) states that only the
hard spectra of AGN could produce enough photons with $h\nu \geq $ 54.4
eV to reionize \ion{He}{2}, based in part on the result $Q_{2}$/$Q_{0}
\leq 0.02$ for low-metallicity stellar spectra to which Wolf-Rayet
stars contribute only minimally (Leitherer \& Heckman~1995).  However,
for the Pop~III cluster in Figure~\ref{fig4}, $Q_{2}$/$Q_{0} = 0.05$.
Thus, for $n_{He} / n_{H} = 0.0785$ ($Y = 0.239$) and assuming full
ionization of H and He, the \ion{He}{3} region excited by this
cluster will have 50\% the radius of its \ion{H}{2} Str\"{o}mgren
sphere.  This large \ion{He}{3} region may imply that \ion{He}{2}
reionization differs in the denser regions, compared to the low-density
IGM where recombinations are not important.

\section{COSMOLOGICAL IMPLICATIONS} 

Pop~III stars can be distinguished from metal-enriched stars by their
theoretical radii and effective temperatures.  Unfortunately, these
features are not observable directly.  However, these characteristics
modify the spectrum of the stars in ways are that are potentially
observable. We discuss these in brief below, but defer detailed studies
of metal-free stars to future work.

If the ionizing spectra of clusters follow a power law, $L_{\nu}
\propto \nu ^{-\alpha}$, we can use $\alpha$ to characterize the
effects of Pop~III stars on the intergalactic radiation field.  The
Pop~III cluster in Figure~\ref{fig4} has a hard (and inverted) spectral
index $\alpha _{1} = - 1.2$ in the range 1 -- 4 Ryd and $\alpha _{4} =
2.0$ for $h\nu \geq 4$ Ryd. For comparison, the Pop~II cluster in
Figure~\ref{fig4} has $\alpha _{1} = 1.0$ and negligible flux above 4
Ryd. Thus, metal-free stars could contribute harder radiation to the
extragalactic spectrum than QSOs with intrinsic $\alpha _{s} = 1.8$
between 0.9 -- 2.6 Ryd (Zheng et al.~1997).  At $z \sim 5$ this hard
EUV spectrum is accessible to ground- and space-based instruments in
the optical and near-UV.

The dramatic increase in \ion{He}{2} ionizing photons from metal-free
stars implies that they will have distinctive effects on their
neighborhoods.  Simple models of Pop~III \ion{H}{2} regions show that
Pop~III stars excite sizable \ion{He}{3} regions.  The $\lambda$1640
and $\lambda$4686 recombination lines of \ion{He}{2} might be detected
for targets with $z = 5 - 10$ in the 1 - 5 $\mu$m range with NGST.
Similarly, the radio recombination lines of He II may provide a unique
signature of these stars. However, \ion{He}{2} recombination emission
observed in the spectra of metal-poor extragalactic \ion{H}{2} regions
has often been attributed to Wolf-Rayet stars, X-ray binaries, or
shocks.  (Garnett et al. 1991; de Mello et al.~1998; Izotov et al.
1997). These sources are likely to be less important at high $z$, but
they may complicate the identification of metal-free stellar
populations.

The photodissociation of H$_{2}$ by FUV radiation (912 -- 1126 \AA)
from the first luminous sources in the universe is suggested to have
inhibited subsequent star formation near these sources by destroying
their only coolant (Haiman, Rees, \& Loeb~1997; Ciardi, Ferrara, \&
Abel~1999).  Photons with $h\nu$ = 11.2 -- 13.6 eV can propagate
freely into neutral, dust-free gas and dissociate H$_{2}$ by exciting
permitted transitions to the $^1 B\, \Sigma _{u} ^{+}$ state, 10 --
15\% of which decay into the continuum.  This ``negative feedback'' is
presumed to precede the ionization front by a distance dependent on the
spectrum of the ionizing source. Pop~III clusters, with enhanced
ionizing photon production and suppressed FUV flux, may dissociate
H$_{2}$ with their ionization fronts.


The hard spectra of Pop~III stars may also affect the IGM ionization
ratios through changes to the extragalactic spectrum.  Fardal et
al.~(1998) define the ratio $\eta \equiv N_{\rm He II} / N_{\rm H I}$
to express the relative column densities of \ion{He}{2} and
\ion{H}{1}.  This ratio is sensitive to the shape of the
extragalactic spectrum and can be used to predict the 
optical depths $\tau _{\rm H I}$ and $\tau _{\rm He II}$ in the IGM. An
increase in the \ion{He}{2} ionizing flux favors \ion{He}{3} and
decreases $\eta$.  Fardal et al.~(1998) estimate that $\eta \gtrsim
100$ is necessary to explain $\tau _{\rm He II} \sim$ 1 -- 5 observed
at $z \sim 3$ (Davidsen, Kriss, \& Zheng~1996).  The $Z = 0$
stellar spectra give $\eta \sim 30$, inconsistent with the
observed opacity if these stars are still forming at that epoch. Thus, 
$Z = 0$ stars may not dominate the intergalactic radiation field at $z
= 3$.

In summary, we outline a general picture of the era of Pop~III stars
based on our models and consistent with the observations discussed
above. We assume that the first stars formed at $z \sim 10$, consistent
with simulations of large-scale structure.  If the total cosmic star
formation rate exceeded the critical rate calculated in \S~4, then
these first stars may have reionized hydrogen and helium in the
universe.  Upon their deaths, they enriched the universe to an average
metallicity 20\% of that observed at $z \sim 3$. Population III then
faded, leaving their metal-enriched progeny to provide the rest of the
metals and the softer radiation seen at $z \sim 3$.

\acknowledgements 

This work was supported in part by astrophysical theory grants from
NASA (NAG5-7262) and NSF (AST96-17073).

\pagebreak 
\centerline{\epsfxsize=\hsize{\epsfbox{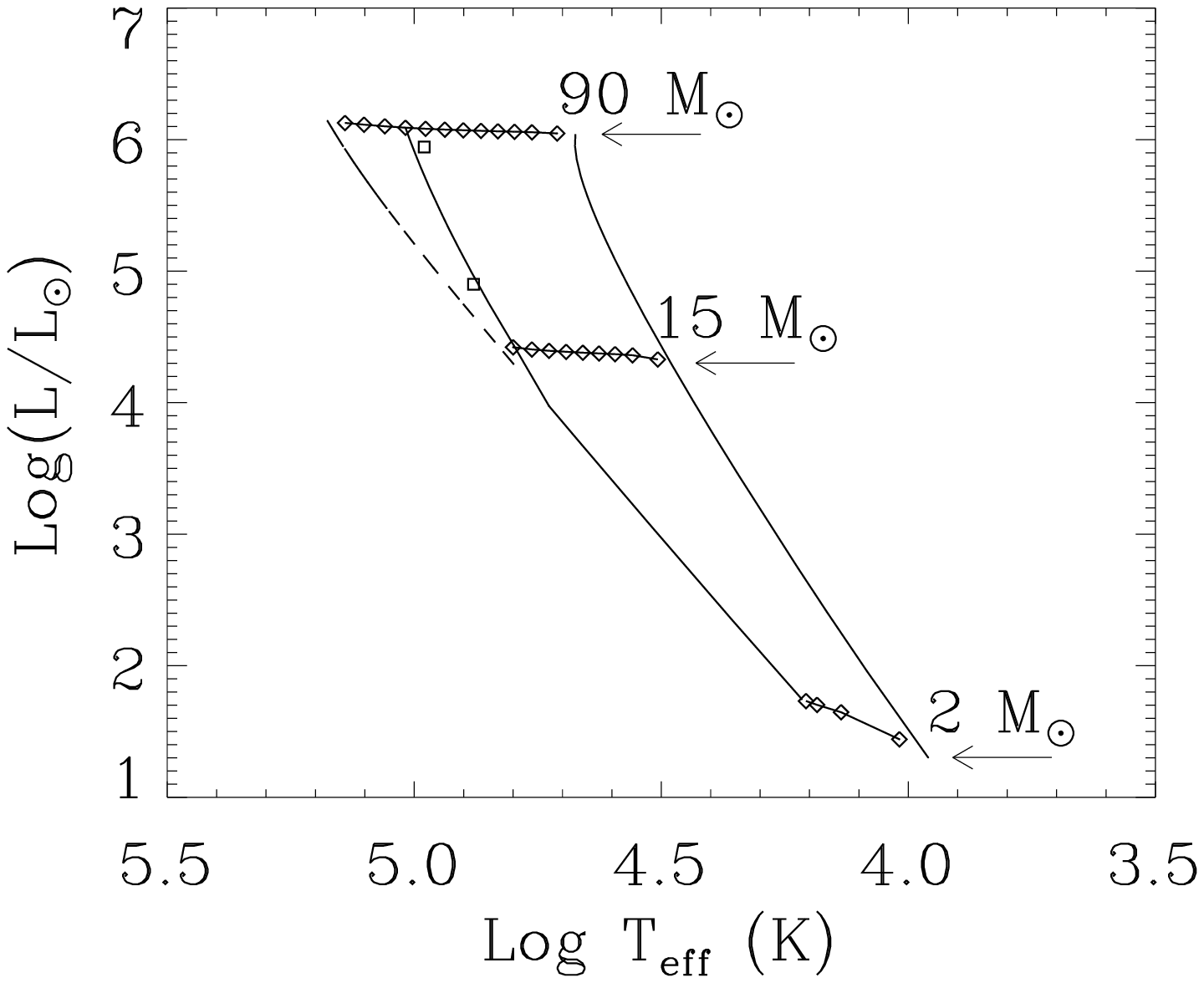}}}
\figcaption{Zero-age main sequences for Pop~I ($Z_{\odot}$ = 0.02) and
Pop~III stars of mass 2, 5, 8, 10, 15, 20, 25, 30, 35, 40, 50, 60, 70,
80, and 90 \Msun. The diamonds mark decades in metallicity in the
approach to $Z = 0$ from 10$^{-2}$ down to 10$^{-5}$ at 2 \Msun, down
to 10$^{-10}$ at 15 \Msun, and down to 10$^{-13}$ at 90 \Msun.  The
dashed line along the Pop~III ZAMS assumes pure H-He composition, while
the solid line marks the upper MS with $Z_{\rm C} =
10^{-10}$ for the $M \geq 15$ \Msun\ models. Squares mark the points
corresponding to pre-enriched evolutionary models from El Eid et al.
(1983) at 80 \Msun\ and from Castellani et al.~(1983) for 25 \Msun.
\label{fig1}}

\centerline{\epsfxsize=\hsize{\epsfbox{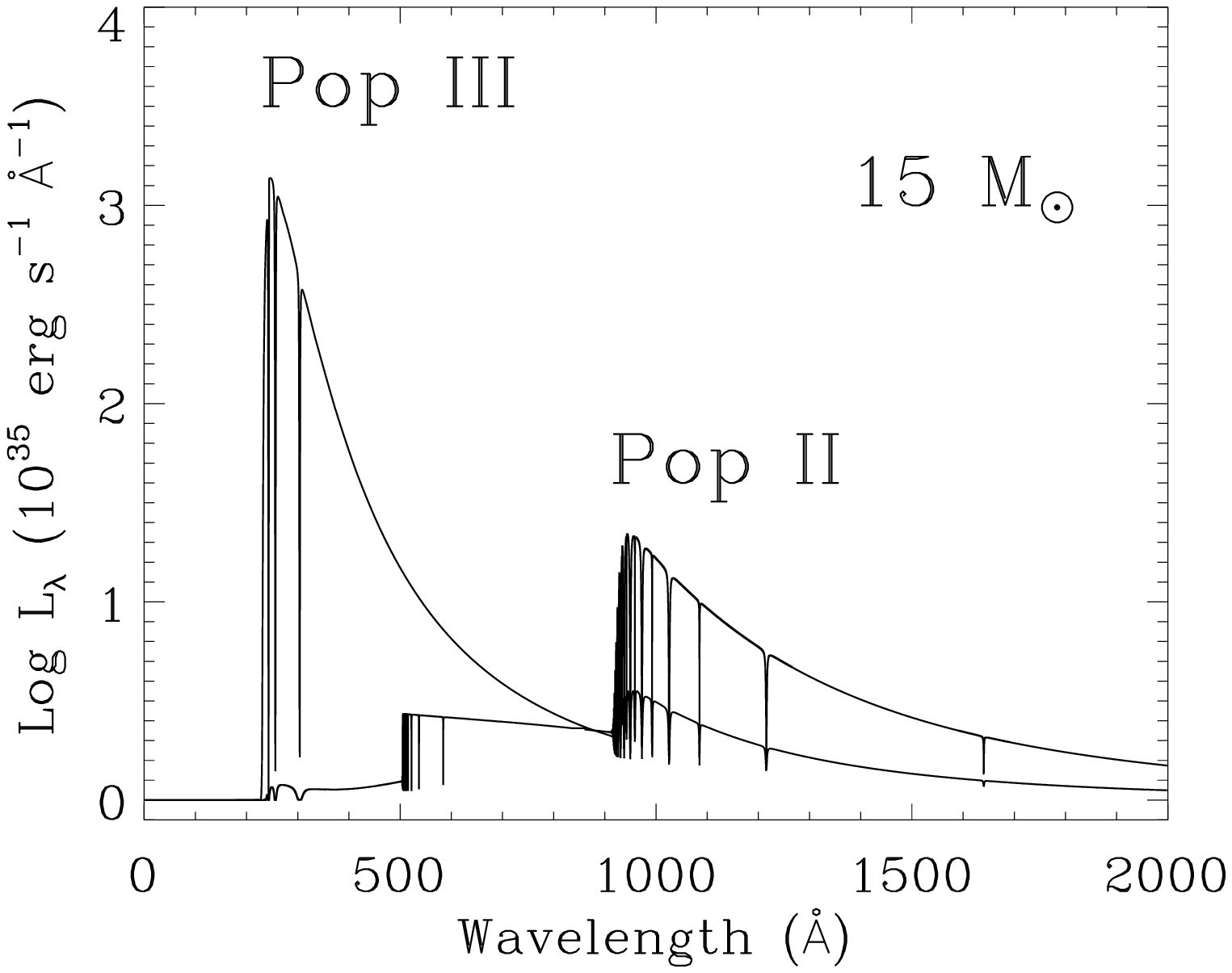}}}
\figcaption{Comparison of spectra calculated from atmosphere models of
Pop~II ($Z = 0.001$) and Pop~III ($Z = 0$) stars of 15 \Msun. The Pop~II
star has $T_{\rm eff}$~=~36,000 K, and the Pop~III star has $T_{\rm
eff}$~=~63,000 K.  Only \ion{H}{1}, \ion{He}{1}, and \ion{He}{2} lines
are included. \label{fig2}}

\centerline{\epsfxsize=\hsize{\epsfbox{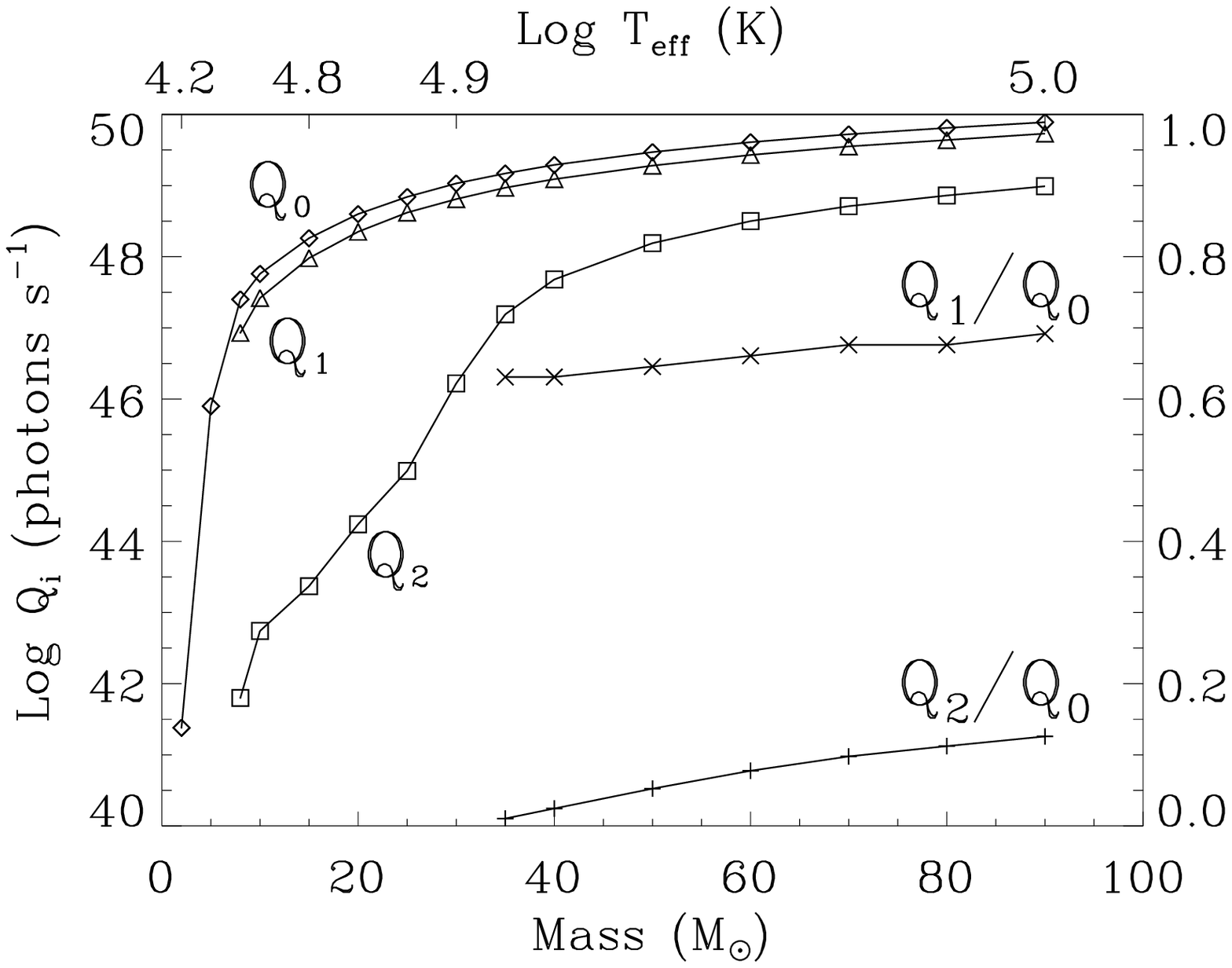}}}
\figcaption{Ionizing photon yields for initially metal-free stars
pre-enriched to $Z_{\rm C} = 10^{-10}$.  High-mass ($M > 25$ \Msun)
stars emit 60 -- 70\% of their ionizing photons in the \ion{He}{1}
ionizing continuum and 2 -- 12\% in the \ion{He}{2} continuum. The
right axis corresponds to the lower two curves, which represent the
ratios $Q_{1}/Q_{0}$ and $Q_{2}/Q_{0}$ of ionizing photons emitted in
the \ion{He}{1} and \ion{He}{2} continua. The top axis is $\log T_{\rm
eff}$ for the displayed models.  \label{fig3}}

\centerline{\epsfxsize=\hsize{\epsfbox{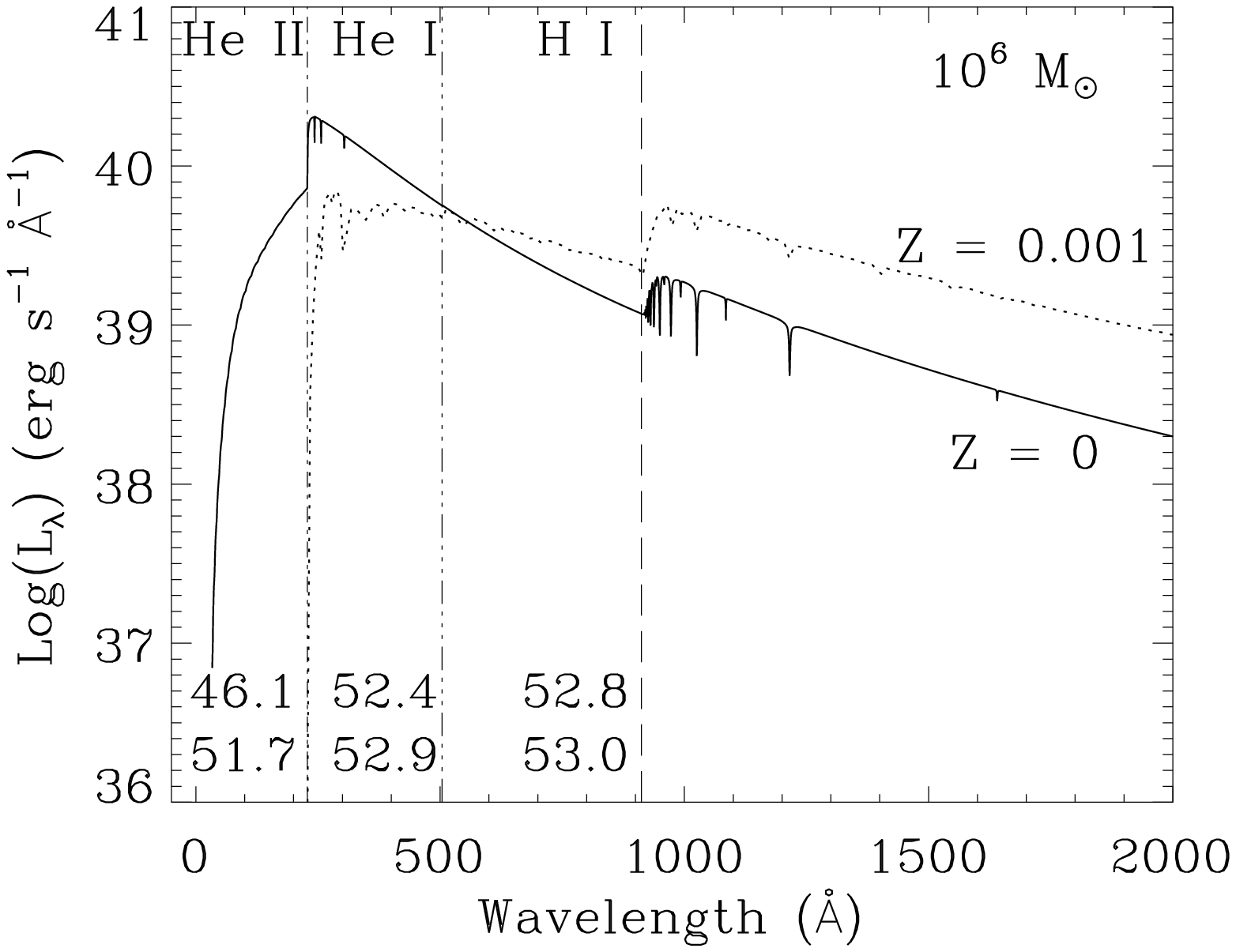}}}
\figcaption{Synthetic spectra of Pop~II and Pop~III clusters of
$10^{6}$ \Msun\ with a Salpeter IMF. The Pop~II spectrum ($Z = 0.001$)
represents an instantaneous burst of star formation at an age of 1 Myr
and is from Leitherer et al.~(1999). The Pop~III cluster was
constructed with the pre-enriched MS displayed in Figure~\ref{fig1}. The
dashed lines mark the three continuum transitions of H I, He I, and He
II, from right to left respectively.  The numbers in the lower left,
near each continuum mark, represent $\log Q_{i}$ for that continuum,
with Pop~II given above.  \label{fig4}}

\pagebreak 






\begin{references} 
\reference{abel}   Abel, T., Anninos, P., Norman, M. L., \& Zhang, Y. 1998,
                     \apj, 508, 518 
\reference{cas}    Castellani, V., Chieffi, A., \& Tornambe, A. 1983, \apj, 272,
                     249
\reference{chi}    Chieffi, A., \& Tornambe, A. 1984, \apj, 287, 745
\reference{h2}     Ciardi, B., Ferrara, A., \& Abel, T. 1999, \apj, in press, 
                     astro-ph/9911137
\reference{dkz}    Davidsen, A. F., Kriss, G. A., \& Zheng, W. 1996, Nature, 380, 47
\reference{dem}    de Mello, D. F., Schaerer, D., Heldmann, J., \& Leitherer, C.
                      1998, \apj, 507, 199 
\reference{eleid}  El Eid, M. F., Fricke, K. J., \& Ober, W. W. 1983, A\&A, 119, 54
\reference{Ez}     Ezer, D. 1961, \apj, 133, 159 
\reference{EzC}    Ezer, D., \& Cameron, A. G. W. 1971, Ap\&SS, 14, 399 
\reference{fgs}    Fardal, M. A., Giroux, M. L., \& Shull, J. M. 1998, \aj, 115,
                     2206
\reference{garn}   Garnett, D. R., Kennicutt, R. C., Jr., Chu, Y.-H., \& Skillman, 
                      E. D. 1991, \apj, 373, 458 
\reference{gs}     Giroux, M. L., \& Shull, J. M. 1997, \aj, 113, 1505 
\reference{Nick}   Gnedin, N. Y., \& Ostriker, J. P. 1997, \apj, 486, 581 
\reference{hm}     Haardt, F., \& Madau, P. 1996, \apj, 461, 20 
\reference{hl}     Haiman, Z., \& Loeb, A. 1997, \apj, 483, 21 
\reference{hrl}    Haiman, Z., Rees, M. J., \& Loeb, A. 1997, \apj, 476, 458
\reference{heap}   Heap, S. R., et al. 1999, \aj, in press, astro-ph/9812429 
\reference{hogan}  Hogan, C. J., Anderson, S. F., \& Rugers, M. H. 1997, \aj, 113,
                     1495
\reference{TLUSTY} Hubeny, I.,  \& Lanz, T. 1995, \apj, 439, 875
\reference{Izotov} Izotov, Y., et al. 1997, \apj, 476, 698 
\reference{K92}    Kurucz, R. 1992, in Stellar Populations of Galaxies, ed. 
                     B. Barbuy \& A. Renzini (Dordrecht: Kluwer) 
\reference{s99}    Leitherer et al. 1999, \apjs, 123, 3 
\reference{lh95}   Leitherer, C., \& Heckman, T. M. 1995, \apjs, 96, 9 
\reference{mad}    Madau, P., Haardt, F., \& Rees, M. J. 1999, \apj, 514, 648  
\reference{ma}     Madau, P., Mozetti, L., \& Dickinson, M. 1998, \apj, 498, 106
\reference{Pei}    Pei, Y.C. 1995, \apj, 438, 623 
\reference{P}      Pettini, M. 1999, in Chemical Evolution
		     from Zero to High Redshift, ed. J. Walsh, \& M.
		     Rosa (Berlin:  Springer), astro-ph/9902173
\reference{Rei}    Reimers, D., Kohler, S., Wisotzki, L., Groote, D., 
                   Rodriguez-Pascual, P., \& Wamsteker, W. 1997, A\&A, 327,
                     890
\reference{OPAL}   Rogers, F. J., \& Iglesias, C. A. 1992, \apjs, 79, 507
\reference{gen}    Schaerer, D., de Koter, A., Schmutz, W., \& Maeder, A. 1996, 
                     A\&A, 312, 475 
\reference{sdk}    Schaerer, D., de Koter, A. 1997, A\&A, 322, 598 
\reference{sch}    Schneider, D., Schmidt, M., \& Gunn, J. E. 1991, \aj, 101,
                     2004
\reference{sc}     Songaila, A., \& Cowie, L. L. 1996, AJ, 112, 335 
\reference{ss}     Sutherland, R. S., \& Shull, J. M. 1999, in preparation 
\reference{me}     Tumlinson, J., Giroux, M. L., Shull, J. M., \& Stocke, J. T. 
                     1999, \aj, 118, in press, astro-ph/9907174 
\reference{vgs}    Vacca, W. D., Garmany, C. D., \& Shull, J. M. 1996, \apj, 460,
                     914
\reference{ww}     Woosley, S. E., \& Weaver, T. A. 1995, \apjs, 101, 181
\reference{Z}      Zheng, W., Kriss, G. A., Telfer, R.C., Grimes, J. P., \& 
                     Davidsen, A. F. 1997, \apj, 475, 469 
\end{references}
\end{document}